\documentclass[sigconf, screen]{acmart}

\acmConference[Arxiv]{}{2025}{Salt Lake City, USA}

\renewcommand\footnotetextcopyrightpermission[1]{} 
\usepackage{amsmath}
\usepackage[lined, boxed, ruled, noend]{algorithm2e}
\usepackage{listings}
\usepackage{xcolor}
\usepackage{subfigure} 
\usepackage{float}
\usepackage{multirow}
\usepackage{booktabs}
\usepackage{array}
\newcolumntype{x}[1]{>{\centering\arraybackslash}m{#1}}
\usepackage{graphicx}
\usepackage{caption}
\usepackage{subcaption}
\usepackage{placeins}  
\usepackage{dblfloatfix}  

\newfloat{FloatLst}{htbp}{loc}
\floatname{FloatLst}{Listing}

\usepackage{marginnote} 
\usepackage{xcolor}

\begin{document}
\title{Is Sparse Matrix Reordering Effective for Sparse Matrix-Vector Multiplication?}

\author{Omid Asudeh}
\orcid{0000-0002-5737-1780}
\affiliation{%
 \institution{University of Utah}
 \streetaddress{}
 \city{Salt Lake City}
 \state{Utah}
 \country{United States}
 \postcode{}
}
\email{asudeh@cs.utah.edu}

\author{Sina Mahdipour Saravani}
\orcid{0000-0003-4285-1439}
\affiliation{%
 \institution{University of Utah}
 \streetaddress{}
 \city{Salt Lake City}
 \state{Utah}
 \country{United States}
 \postcode{}
}
\email{sina@cs.utah.edu}

\author{Fabrice Rastello}
\orcid{0000-0002-6589-9956}
\affiliation{%
 \institution{INRIA}
 \city{Grenoble}
 \country{France}
}
\email{fabrice.rastello@inria.fr}

\author{Gerald Sabin}
\orcid{0000-0002-8672-4071}
\affiliation{%
 \institution{RNET Technologies}
 \city{Dayton}
 \state{Ohio}
 \country{United States}
}

\email{gsabin@rnet.com}

\author{P. Sadayappan}
\orcid{0000-0002-4737-2034}
\affiliation{%
 \institution{University of Utah}
 \city{Salt Lake City}
 \state{Utah}
 \country{United States}
}
\email{saday@cs.utah.edu}

\begin{abstract}
This work evaluates the impact of sparse matrix reordering on the performance of sparse matrix-vector multiplication across different multicore CPU platforms. Reordering can significantly enhance performance by optimizing the non-zero element patterns to reduce total data movement and improve the load-balancing. We examine how these gains vary over different CPUs for different reordering strategies, focusing on both sequential and parallel execution. We address multiple aspects, including appropriate measurement methodology, comparison across different kinds of reordering strategies, consistency across machines, and impact of load imbalance.
\end{abstract}
\maketitle

\section{Introduction}
\label{sec:intro}

Sparse Matrix-Vector Multiplication (SpMV) is a fundamental computational kernel in numerous scientific and engineering applications. Unlike its dense counterpart, optimizing SpMV is significantly more challenging due to the inherent sparsity of the matrix. This sparsity can lead to load imbalance across threads, induce irregular memory access patterns, 
and reduces data reuse, thereby exacerbating data movement bottlenecks.

The sparsity structure of a matrix varies greatly depending on the application domain and the nature of the data it represents. This structure plays a crucial role in shaping the performance characteristics of SpMV. In particular, it influences the extent of load imbalance, memory access irregularity, and the pressure on memory bandwidth. 

\begin{figure*}[!ht]
\centering
\begin{minipage}{.46\textwidth}
\centering
\includegraphics[width=\linewidth]{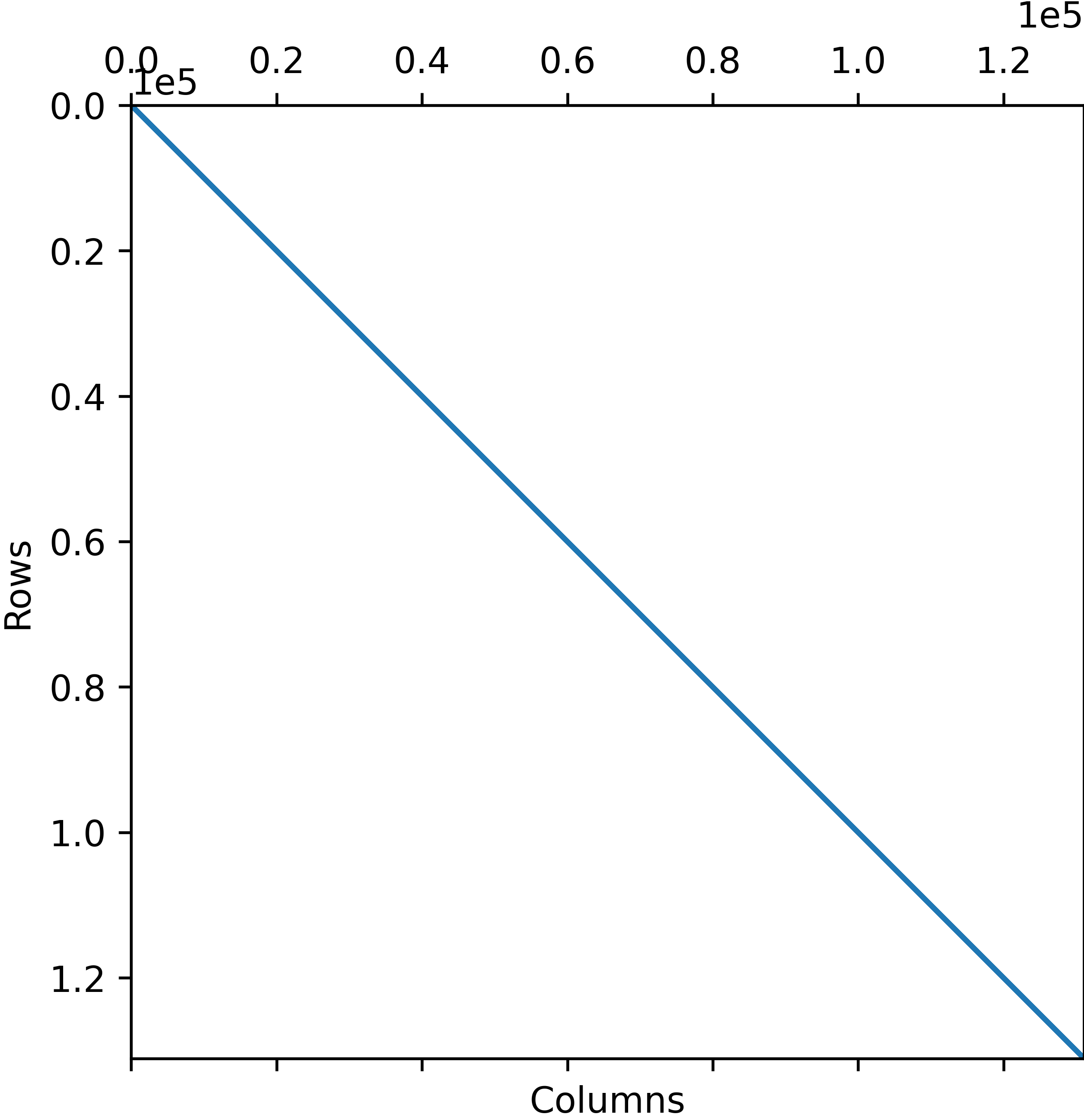}
\caption*{(a) Original banded matrix}
\label{fig:mat_banded}
\end{minipage}
\hspace{-0.5em}
\begin{minipage}{.495\textwidth}
\centering
\includegraphics[width=\linewidth]{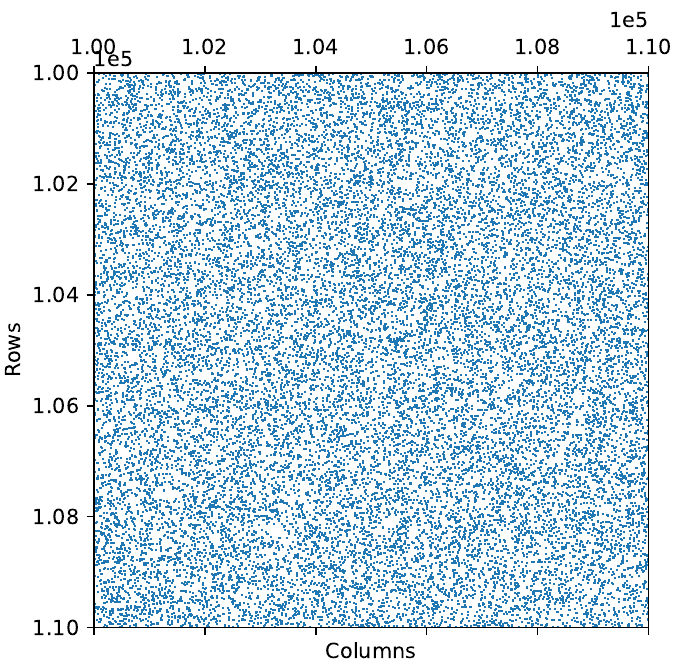}
\caption*{(b) Shuffled version (zoomed)}
\label{fig:mat_shuffled}
\end{minipage}
\caption{Comparison of an original banded matrix of size $128\mathrm{K} \times 128\mathrm{K}$ with band size 15, and its shuffled counterpart (zoomed).}

\label{fig:banded}
\end{figure*}

We illustrate the impact of the sparsity structure on performance via a simple synthetic experiment. Starting with a banded matrix ($128K \times 128K$, bandwidth $63$), we perform a random symmetric row/column permutation to create a sparse matrix with uniformly random distribution of the non-zeros in the index space, as shown in Fig.~\ref{fig:banded}. Using a CSR representation for both the matrices, the measured performance of SpMV on a 64 core multiprocessor for the two matrices is very different: 108 GFLOPs for the former and 32 GFLOPs for the latter.


Consequently, considerable effort has been dedicated to developing reordering techniques that permute the rows and columns of a sparse matrix to alter its structural properties with the goal of improving performance. 
Matrix reordering — often approached through graph or hypergraph reindexing — has been extensively explored for SpMV on multicore architectures~\cite{Azad2016TheRC, Rolinger2018ImpactOT, Gao2024ASL, sc23}. However, despite numerous contributions, there is no consensus on consistent use or best practices for reordering in the shared-memory setting. In contrast, for distributed-memory systems, hypergraph partitioning techniques have been widely adopted and shown to significantly improve SpMV performance~\cite{Slota2017DistributedGL, Akbudak2012TechnicalRO, Drescher2023BOBAAP, sc23, Gao2024ASL, Sun2018VEBOAV, Aupy2016LocalityAwareLM, Munson2005TheFB}.

We consider two types of deployment scenarios for sparse matrix-vector multiplication applications: i) executing a parallel SpMV implementation on a multicore processor, and ii) executing multiple independent instances of sequential implementation of SpMV across cores, where the parallel application inherently has ample outer-level parallelism (e.g., multiple concurrently solved systems of equations, where each uses sequential SpMV). Therefore, we compare the the impact of reordering methods on the performance of both parallel and sequential SpMV.

In this paper, we present a comprehensive experimental study of matrix reordering for SpMV on multicore systems, using a large and diverse set of matrices from the SuiteSparse~\cite{suitesparse} collection. We seek to answer the following research questions:
\begin{enumerate}
\item   Does the measurement methodology affect the observed performance?
\item Are some reordering schemes consistently effective across matrices?
\item Do reorderings show consistent behavior across different machines?
\item Does matrix reordering influence load balancing?
\end{enumerate}

\section{Background and Experimental Setup}
\label{sec:background}
\subsection{Sparse Matrix Reordering Schemes}
To explore the performance impact of reordering on sparse matrix computations, we consider a diverse set of reordering schemes. These were selected to represent a broad spectrum of strategies, including graph-based, hypergraph-based, spectral, and community-based approaches. By covering these distinct paradigms, we aim to capture the properties that affect memory locality, parallelism, and load balancing in sparse matrix kernels.


\textit{METIS Graph Partitioning.}  
METIS is a multilevel graph partitioning algorithm that recursively coarsens the input graph, partitions the coarsest version, and then refines the partitioning during uncoarsening. It aims to minimize edge cuts while balancing partition sizes. When applied to sparse matrices, METIS can enhance cache locality and reduce fill-in during matrix factorizations by clustering highly connected nodes together ~\cite{karypis1997metis, karypis1998fast}.


\textit{PaToH Hypergraph Partitioning.}  
PaToH is a multilevel hypergraph partitioning tool optimized for irregular data structures. It models multi-way relationships more effectively than standard graphs and improves data locality by minimizing cut hyperedges and balancing workload across partitions, which benefits operations like sparse matrix-vector multiplication ~\cite{atalyrek1999HypergraphPartitioningBasedDF}.


\textit{RCM (Reverse Cuthill–McKee) Ordering.}  
RCM aims to reduce the bandwidth of sparse matrices. It performs a breadth-first traversal starting from a low-degree node and visits neighboring nodes in order of increasing degree. The resulting ordering is reversed, typically achieving better bandwidth reduction than the standard Cuthill–McKee method and improving memory locality during computations ~\cite{Gibbs1976AnAF, Liu1976ComparativeAO}.


\textit{Louvain Community Detection.}  
Louvain method is a hierarchical clustering algorithm that detects communities in a graph by maximizing modularity. It first identifies small communities and then aggregates them into super-nodes, repeating the process. As a matrix reordering technique, it clusters strongly connected nodes together, which can improve performance by exploiting community structure in the graph ~\cite{Blondel2008FastUO}.

\subsection{Experimental Setup}

To evaluate the performance impact of various reordering schemes, we conducted experiments on multiple modern CPUs using carefully controlled runtime settings. Below, we describe the hardware configurations, software environment, runtime settings, and dataset used in our experiments.

\noindent\textit{\textbf{Hardware Platforms}}

Our experiments were conducted on a diverse set of CPU architectures to evaluate the performance impact of reordering across different hardware configurations. The systems used are:

\begin{itemize}
  \item \textbf{AMD-Server:} AMD Ryzen Threadripper 3990X, 64 cores, 128 threads, 256~MiB L3 cache.
  \item \textbf{Intel-Server:} Intel Core i9-10980XE @ 3.00\,GHz, 18 cores, 36 threads, 24.8~MiB L3 cache.
  \item \textbf{Intel-Desktop:} Intel Core i7-11700KF (11th Gen) @ 3.60\,GHz, 8 cores, 16 threads, 16~MiB L3 cache.
  \item \textbf{AMD-Desktop:} AMD Ryzen 7 3700X, 8 cores, 16 threads, 32~MiB L3 cache.
\end{itemize}

These systems span high-core-count servers and consumer desktop CPUs, providing a broad view of performance behaviors under different core counts, memory hierarchies, and vectorization capabilities.

\noindent\textit{\textbf{OS and Compiler}}

All experiments were performed on 64-bit Linux systems using GCC 11.4 as the compiler. 

\noindent\textit{\textbf{Execution Environment}}

To ensure reproducibility and minimize variability due to hardware-level behaviors, the following settings were applied:

\begin{itemize}
  \item \textbf{Hyper-threading/SMT:} Disabled in BIOS across all platforms to ensure threads are mapped to physical cores only.
  \item \textbf{Thread Pinning:} Threads were pinned to physical cores using \texttt{OMP\_PLACES=cores} and \texttt{OMP\_PROC\_BIND=close}.
  \item \textbf{Frequency Control:} Turbo boosting was disabled in BIOS to eliminate frequency-induced performance variance.
  \item \textbf{NUMA Binding:} For NUMA systems, memory allocation was restricted to a single NUMA, ensuring memory access locality.
  \item \textbf{Thread Counts:} Experiments were run both for sequential and parallel execution. For the parallel case, 
 experiments were run with $Cores-1$ OpenMP threads, where $Cores$ denotes the number of physical cores. Sequential execution represents SpMV being used with an outer level of parallelism, and the parallel execution represents a shared memory parallel execution of the SpMV where there is not an outer level of parallelism. 
\end{itemize}

\noindent\textit{\textbf{Dataset}}

\begin{figure}[h!]
  \centering
  \includegraphics[width=0.4\textwidth]{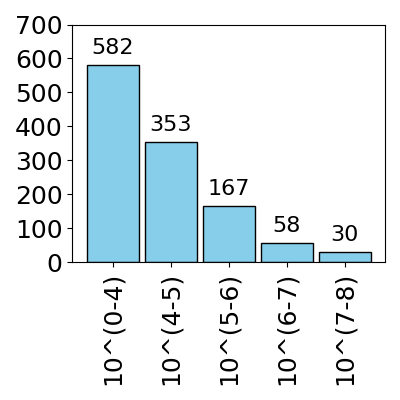}
\caption{Count of symmetric SuiteSparse matrices grouped by matrix size ($m$).}

  \label{fig:matrix-binning}
\end{figure}

The evaluation utilizes a collection of structurally diverse, real-world sparse matrices from the SuiteSparse Matrix Collection~\cite{davis2011university, suitesparse}. A total of 588 matrices were selected, based on two criteria:
\begin{itemize}
  \item \textbf{Symmetric:} We only used symmetric matrices from the collection because some of the reordering algorithms require it, e.g., METIS.
  \item \textbf{Large-scale:} We selected matrices containing more than 10,000 rows, eliminating small matrices because they exhibit high variance in execution time across repeated runs.
\end{itemize}
Figure~\ref{fig:matrix-binning} presents a histogram illustrating the distribution of matrix sizes as a function of the number of rows ($M$). After filtering out matrices that are asymmetric or small, a total of 588 matrices from the SuiteSparse collection remain and are employed in the experiments.

\section{Does Measurement Methodology Matter?}\label{subsec:rq0}
This section examines the influence of two critical components of the benchmarking methodology: the approach to collecting repetitive timing measurements and the choice of OpenMP scheduling scheme. 

\subsection{Approach for Repeated Measurements}
In order to reduce system noise, multiple experiments are performed with each experiment being timed. Caching effects complicate the analysis of repeated computation of Sparse Matrix-Vector Multiplication (SPMV). Even though repeated measurements are intended to reduce noise and increase the reliability of the data collected, the
repeated measurements (Listing~\ref{lst:rax}) often lead to inflated performance due to unnatural data reuse, particularly in parallel algorithms where subsets of input data fit entirely within local L1 or L2 caches. For example, in a simple Compressed Sparse Row (CSR) SPMV, the input vector $x$ may exhibit frequent data reuse. Each row of $A$ accesses multiple elements of $x$, potentially causing a subset of $x$ to remain cached in local L1 or L2 memory during computation (i.e., $x$ is never invalidated or written). For some matrices the subset of X accessed by a single processor may remain resident in local caches between each measured SPMV iteration, exacerbating this effect.
However, real-world applications such as Conjugate Gradient (CG) perform additional tensor operations with a wider variation of per process access patterns requiring data to be moved between local caches. For instance, in CG (Listing~\ref{lst:cg}), each iteration updates vectors used in the next iteration (e.g., the CG solution $x$, residual $r$, and direction $p$). Specifically, the parallel write to $p$ causes a block distribution of $p$ across the processors that does not match the sparse reuse access pattern in the SpMV.

Some prior work attempts to mitigate this by flushing the cache between runs, while others, like Trotter et al.~\cite{sc23}, simply adopt the repeated measurement approach.

To better approximate such behavior in a controlled setting, we propose an alternative repeated measurement approach called input-output swapping (\texttt{IOS}), where the output vector is swapped with the input for the next iteration (Listing~\ref{lst:ios}). This technique better reflects the cache reuse characteristics of real applications where the input vector cannot be cache between uses as readily, while preserving simplicity and benchmark reproducibility.

\lstset{
  basicstyle=\footnotesize\ttfamily,
  breaklines=false,
  keepspaces=true,
  numbers=none,
  columns=flexible
}

\begin{FloatLst}
\begin{lstlisting}[language=C++, caption={RAX: Repeated measurement of y=Ax}, label={lst:rax}]
for (i = 0; i < iters; i++) {
    start_time = omp_get_wtime();
    csr_mv(A_CSR, m, x, y);               // Perform SpMV
    end_time = omp_get_wtime();
    durations[i] = 1000 * (end_time - start_time); // Duration in ms
}
\end{lstlisting}
\end{FloatLst}



\begin{FloatLst}
\begin{lstlisting}[language=C++, caption={IOS: output becomes input to emulate real usage and disrupt cache reuse.}, label={lst:ios}]
for (i = 0; i < iters; i++) {
    start_time = omp_get_wtime();
    csr_mv(A_CSR, m, x, y);
    end_time = omp_get_wtime();
    durations[i] = 1000 * (end_time - start_time);

    temp = x;                            // Swap input/output
    x = y;
    y = temp;
}
\end{lstlisting}
\end{FloatLst}

\begin{FloatLst}
\begin{lstlisting}[language=C++, caption={CG kernel loop highlighting realistic SPMV reuse behavior.}, label={lst:cg}]
for (iter = 0; iter < max_iter; iter++) {
    start_time = omp_get_wtime();
    csr_mv(A_CSR, m, p, Ap);             // Ap = A * p
    end_time_spmv = omp_get_wtime();

    pAp = dot(p, Ap, m);
    alpha = rs_old / pAp;

    #pragma omp parallel for
    for (i = 0; i < m; i++) {
        x[i] += alpha * p[i];            // Update solution
        r[i] -= alpha * Ap[i];           // Update residual
    }

    rs_new = dot(r, r, m);
    beta = rs_new / rs_old;

    #pragma omp parallel for
    for (i = 0; i < m; i++) {
        p[i] = r[i] + beta * p[i];       // Update direction
    }

    rs_old = rs_new;
}
\end{lstlisting}
\end{FloatLst}

\begin{figure}[h]
  \centering
  \includegraphics[width=0.499\textwidth]{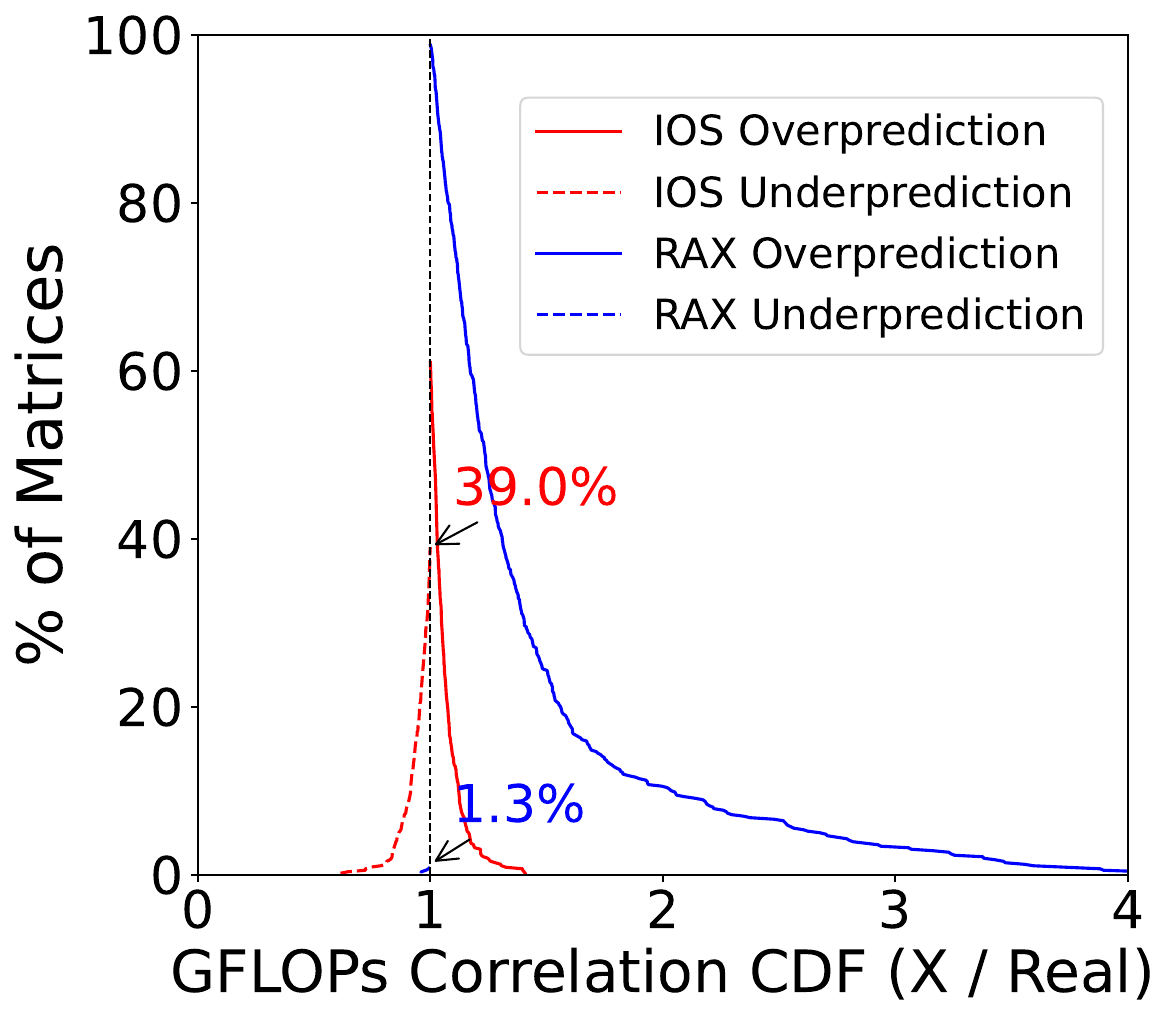}
\caption{CDF plot comparing the performance correlation of \texttt{RAX} and \texttt{IOS} benchmarking methodologies with real application data captured using \texttt{CG}. The plot demonstrates that \texttt{IOS} produces a more gradual and balanced error distribution across matrices, while \texttt{RAX} consistently overestimates performance relative to \texttt{CG}.}

  \label{fig:RQ0_GFLOPS_corr}
\end{figure}

Figure~\ref{fig:RQ0_GFLOPS_corr} presents the cumulative distribution function (CDF) of the ratio between measured GFLOPs by each benchmarking method and the GFLOPs observed in the real application, as captured using the Conjugate Gradient (CG) method. Results are aggregated across multiple machines. The plot compares two benchmarking methodologies: \texttt{RAX}, which conducts repeated measurements without modifying inputs, and \texttt{IOS}, which swaps input and output vectors to more accurately emulate real application dataflow.
The x-axis shows the ratio of the measured performance, $X$, to the real performance, $\texttt{Real}$, where $X \in {\texttt{RAX}, \texttt{IOS}}$, and the y-axis shows the percentage of matrices for which the error exceeds a given threshold. The plot shows that \texttt{IOS} yields a more balanced and gradual error distribution, while \texttt{RAX} is clearly overpredicting the real application. Unless otherwise specified, experiments performed in this paper use the \texttt{IOS} methodology. 

\subsection{\textbf{Impact of OpenMP Scheduling}}


\begin{figure}[ht]
  \centering
  \includegraphics[width=.99\linewidth]{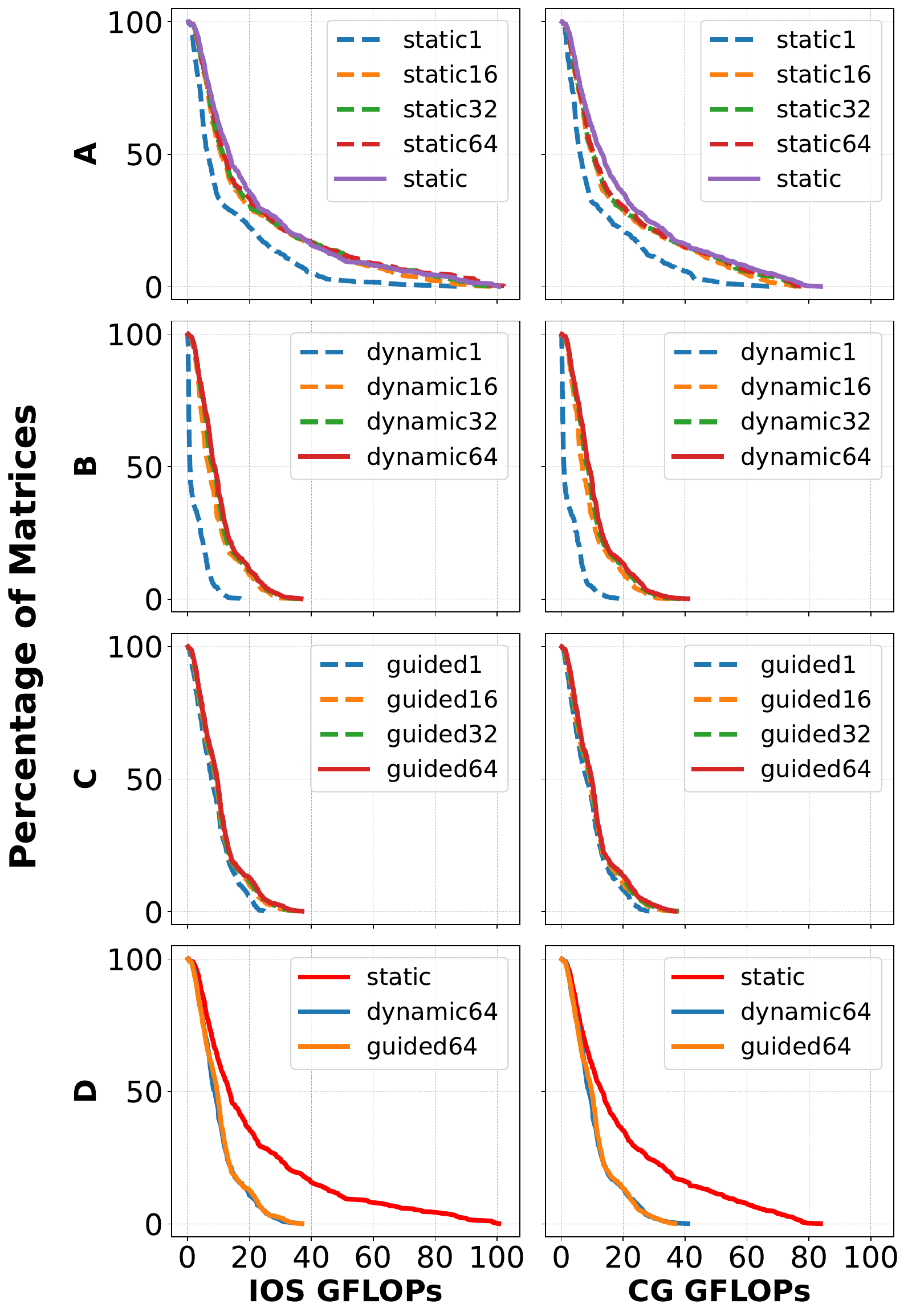}
\caption{Performance of SpMV under different OpenMP schedulings: (A) static, (B) dynamic, (C) guided, and (D) the best configuration selected from static, dynamic, and guided.}
\label{fig:schedules}
\vspace*{-4ex}
\end{figure}

To evaluate SpMV performance in a practical setting, we benchmarked its implementation within the CG algorithm and using IOS, employing various OpenMP scheduling policies: \texttt{static}, \texttt{dynamic}, and \texttt{guided}. These scheduling strategies control how loop iterations, such as matrix rows, are distributed among threads during parallel execution. As illustrated in our \texttt{csr\_mv} kernel (Listing~\ref{lst:csr_spmv}), each row of the sparse matrix is processed independently by iterating over its nonzero elements and computing their dot product with the input vector. The outer loop over matrix rows is parallelized using the OpenMP directive \texttt{\#pragma omp parallel for}.

In \texttt{static} scheduling, OpenMP pre-assigns contiguous blocks of rows to threads before execution, thereby minimizing runtime overhead. This approach assumes a uniform workload across rows, which may not hold for irregular datasets. When a chunk size (also referred to as the \emph{band size}) is specified, rows are cyclically allocated to threads in round-robin fashion, with each thread processing the specified number of rows per cycle. For example, using a static schedule with a chunk size of 16 across 4 threads, thread 0 is assigned rows \{0--15, 64--79, ...\}, thread 1 processes rows \{16--31, 80--95, ...\}, and so forth. In contrast, \texttt{dynamic} scheduling allocates fixed sized chunks of rows to threads at runtime as threads complete their currently assigned chunk, enabling better load balancing for irregular workloads but introducing additional synchronization overhead. \texttt{Guided} scheduling, another dynamic approach, begins by assigning large chunk sizes to threads and progressively reduces the chunk sizes in an exponential manner as the workload diminishes, striking a balance between load distribution and synchronization costs.

We evaluated all three scheduling policies with chunk sizes of \{1, 16, 32, 64\}, as well as the default \texttt{static} schedule, which does not explicitly specify a chunk size. The default \texttt{static} schedule assigns a single, equal-sized chunk of maximal size to each process. Figure~\ref{fig:schedules} presents the cumulative distribution function (CDF) of the observed GFLOPs across matrices, demonstrating that the default \texttt{static} schedule consistently achieves the best performance. Among the static strategies, scheduling overhead is minimized, though at the potential cost of increased load imbalance. Moreover, temporal cache locality improves as the chunk size increases, with the default \texttt{static} schedule benefiting from the maximal chunk size. This relationship is evident from the observed increase in throughput as the chunk size grows. Consequently, we adopt the default \texttt{static} schedule for all subsequent experiments in this paper, unless otherwise specified.
\begin{FloatLst}
\begin{lstlisting}[language=C++, caption={Parallel CSR-based SpMV $y = Ax$ where $A$ is a sparse matrix stored in Compressed Sparse Row (CSR) format.}, label={lst:csr_spmv}]
void csr_mv(A_CSR,m,x,y)
{
    #pragma omp parallel for
    for (uint32_t r = 0; r < m; r++) 
    {
        COMPUTETYPE sum = 0;
        for (uint32_t idx = A_CSR->rowPtr[r]; 
             idx < A_CSR->rowPtr[r + 1]; 
             idx++) 
        {
            uint32_t col = A_CSR->cols[idx];
            COMPUTETYPE val = A_CSR->values[idx];
            sum += val * x[col];
        }
        y[r] = sum;
    }
}
\end{lstlisting}
\end{FloatLst}

 \section{Is any reordering scheme consistently superior?}\label{subsec:rq1}




\begin{figure}[t]
  \centering
  \includegraphics[width=0.9\linewidth]{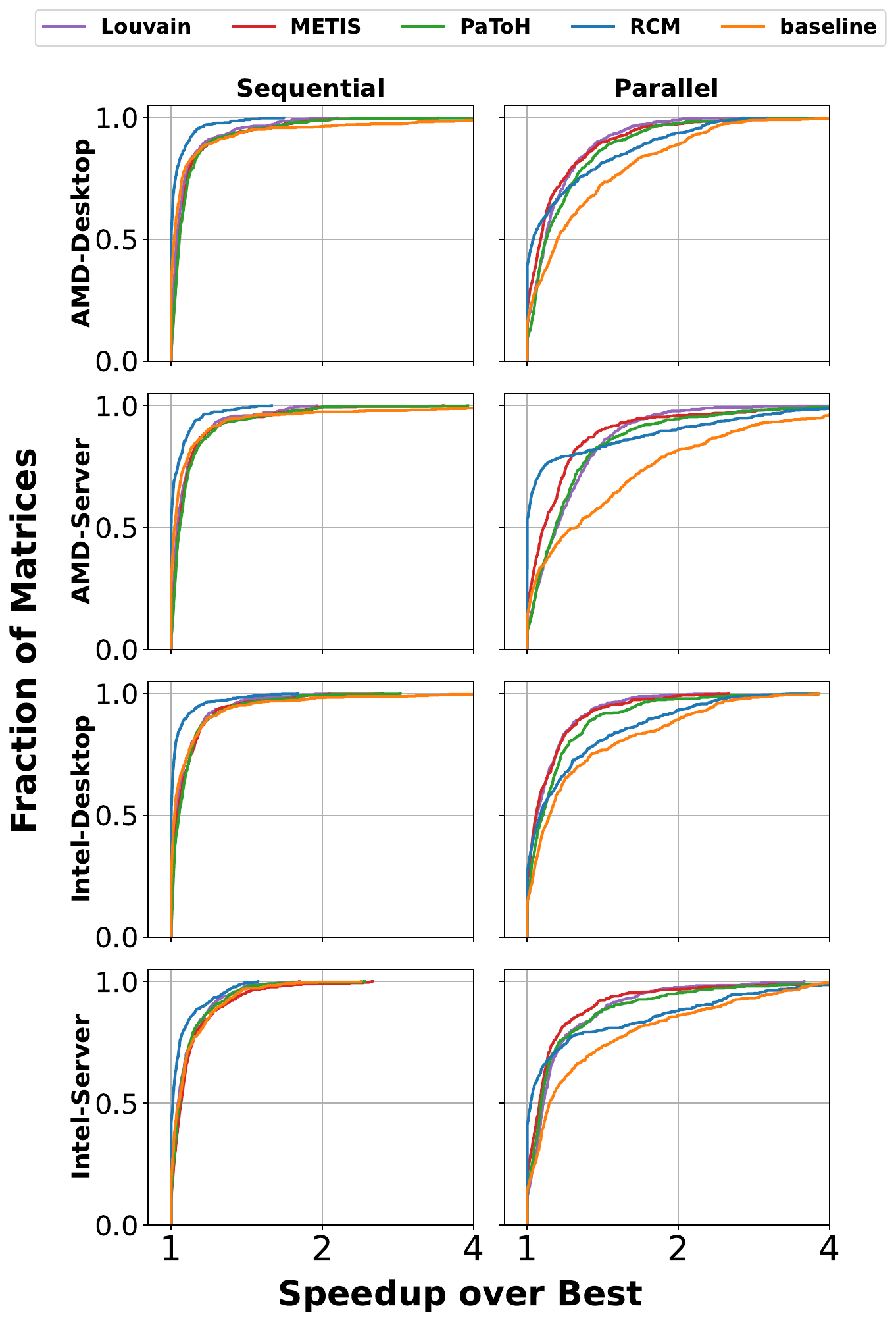}
  \caption{Performance profile plots of different reorderings}
  \label{fig:RQ1_perf_profiles}
  \vspace*{-4ex}
\end{figure}

\subsection{\textbf{Performance Benefits of Reordering}}
As discussed in Section~\ref{subsec:rq0}, we compare different reordering schemes with Input Output Swap (IOS) measurements to draw conclusions that correlate more accurately with real-world applications. In the parallel case, we present measurement results using the default OpenMP static schedule.

Instead of simply reporting observed absolute performance before and after reordering or showing box-plots of the distribution of speedup from reordering, we found it more insightful to perform comparisons across the reordering schemes to use
performance profile plots ~\cite{dolan2002benchmarking}. Figure~\ref{fig:RQ1_perf_profiles} compares the speedup achieved for each reordering scheme on all matrices. Each curve shows the performance profile of a reordering method. The performance profile plot for each performance ratio $\tau>=1$ shows the fraction of matrices for which the corresponding reordering scheme achieves a performance within $\tau$~times the  performance from the best reordering scheme for that matrix. 
For sequential execution, as observed in the top row of the grid, RCM consistently outperforms other schemes on three of the four platforms and is among the best for the fourth (Intel-Server). 


\begin{figure}[ht]
  \centering
  \includegraphics[width=.99\linewidth]{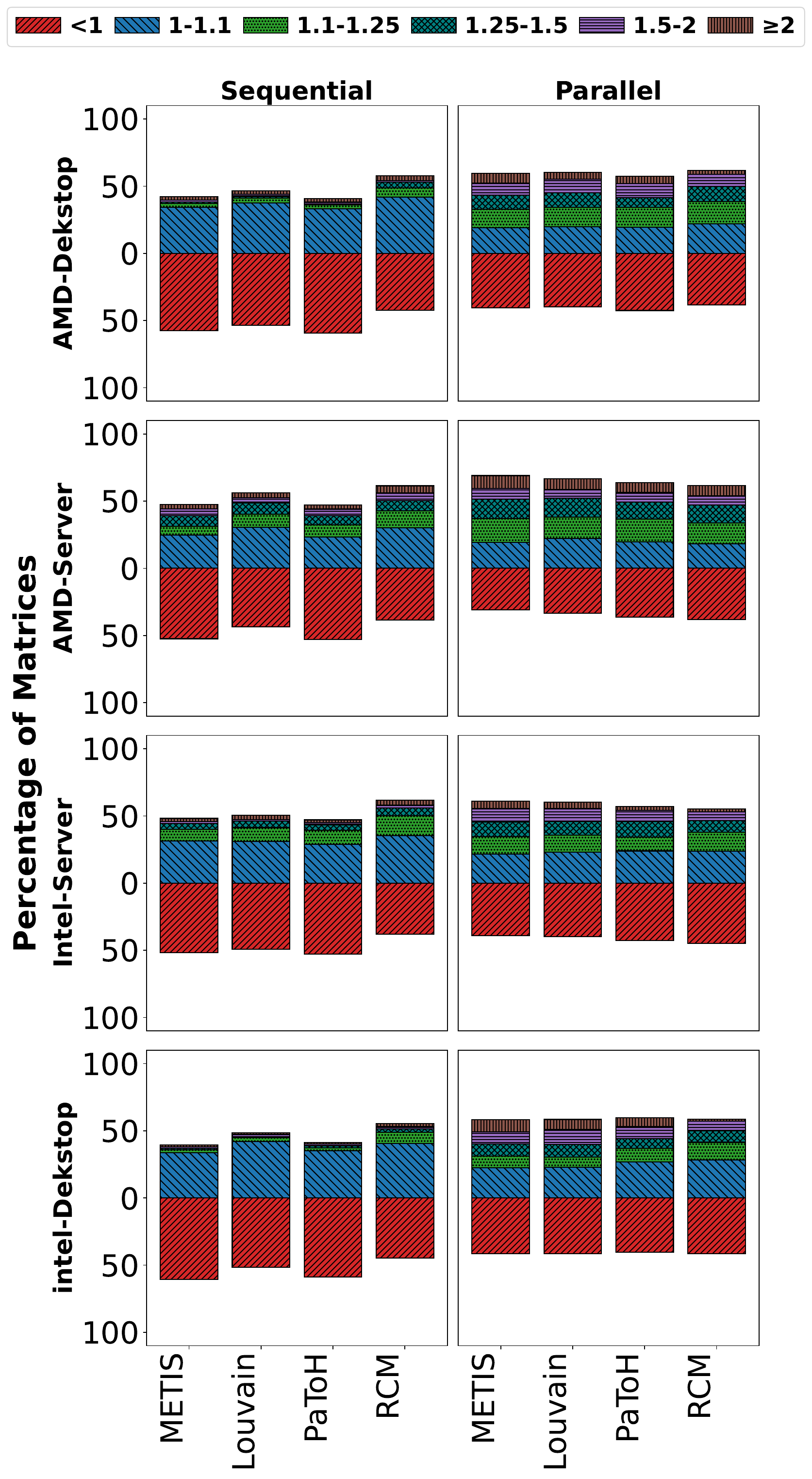}
\caption{Speedup/slowdown distribution for all reordering schemes and machines for sequential and parallel settings.}
\label{fig:stack}
\vspace*{-4ex}
\end{figure}

In the parallel case (second row of the grid), however, the conclusion is not dominantly in favor of either reordering scheme. RCM has an edge in lower threshold $\tau$ values, 
while METIS improves the performance for higher fraction of matrices if considering a higher value of $\tau$. This implies that RCM has the best performance for the most number of matrices, but METIS is able to achieve a higher speedup in many cases.  

In Figure \ref{fig:RQ1_perf_profiles} it can be observed that there is not a consistent trend that can be observed from the performance profiles.




\begin{figure}[ht]
  \centering
  \includegraphics[width=0.99\linewidth]{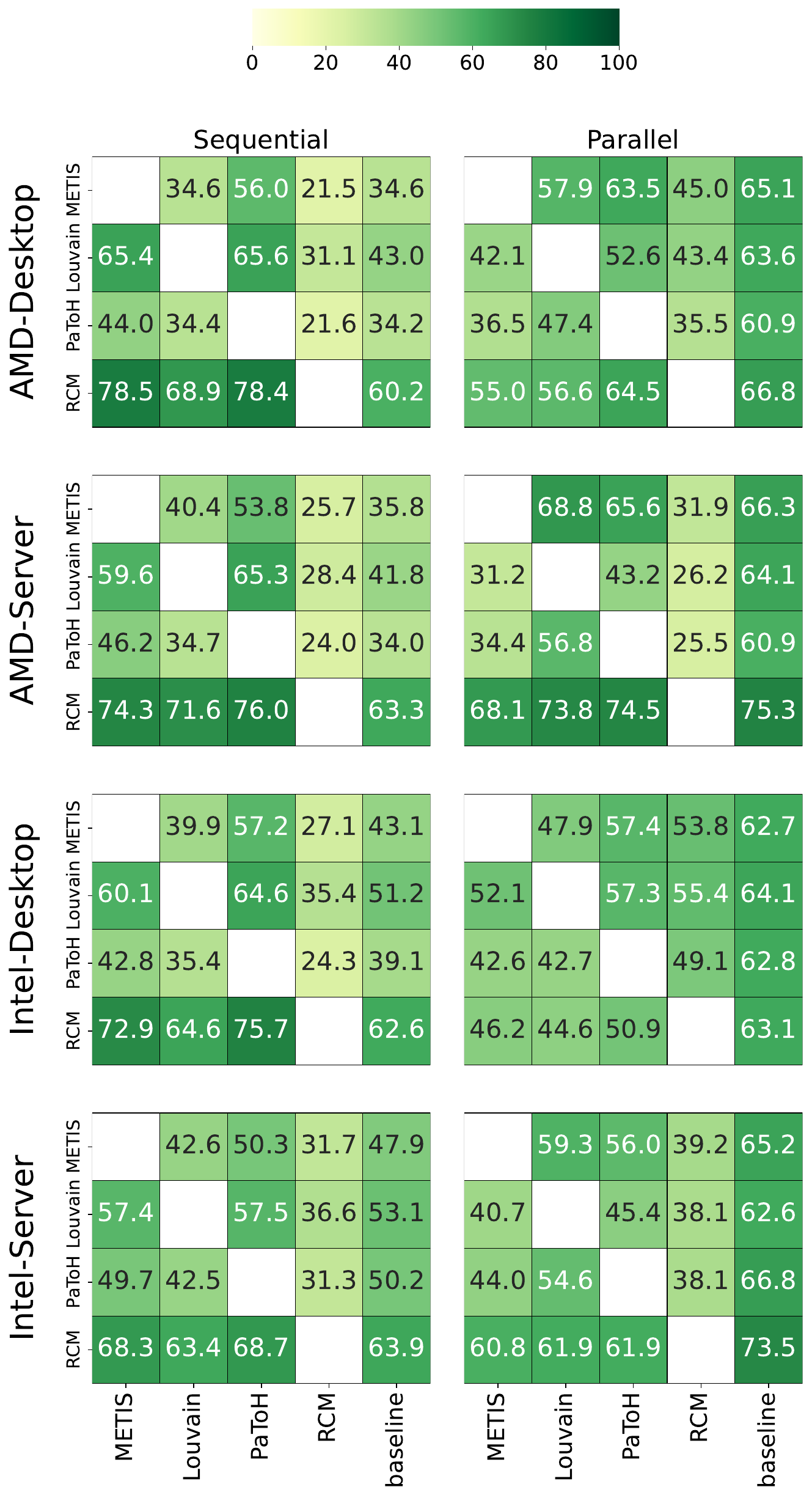}
  \caption{Comparison of reordering GFLOPS in both sequential and parallel settings. Cell values show win rate: the fraction of matrices for which the reordering corresponding to the row-label outperforms the reordering scheme corresponding to the column label ($win\_rate=\frac{|improved\_matrices|}{|all\_matrices|}$).}
  \label{fig:winrate_heatmap}
\end{figure}

\subsection{\textbf{Matrix Speedup from Reordering}}
While Figure~\ref{fig:RQ1_perf_profiles} presents an overall relative comparison of the reordering methods, it does not provide any standalone insight into how well each reordering performs in terms of the actual extent of performance improvement achieved via reordering. 


The stacked bar charts in Figure~\ref{fig:stack} summarize the impact of various reordering strategies on performance across multiple CPU architectures, under both \textbf{sequential and parallel} execution. Each group of bars corresponds to a specific hardware platform, with the top row showing sequential results and the bottom row showing parallel results. Within each group, bars represent different reordering methods, and the colors indicate the number of matrices falling into specific speedup ranges (e.g., \textless1, 1–1.1, 1.1–1.25, etc.). This visualization provides a clear breakdown of how many matrices benefited from each reordering strategy and by how much, including cases of slowdown.
In the sequential case, surprisingly more than 50\% of the matrices experience a slowdown with every reordering strategy except RCM. Interestingly, even in the parallel case, a non-trivial proportion of matrices still experience a slowdown with most reordering strategies, though the impact is generally less severe than in the sequential setting. This behavior is somewhat counterintuitive, as reordering is typically expected to improve performance or at least not hurt performance in the majority of cases.


A clear trend that emerges is that RCM performs best in the sequential setting, achieving the highest number of matrices with speedups no matter the threshold that is considered. However, its advantage does not simply carry over to the parallel case. In the parallel execution, while by considering a pure speedup -- threshold equal to 1 -- RCM still proves as the most effective, METIS shows consistently strong performance with larger threshold values.
This is specifically interesting as what is conveyed is that while RCM provides a speedup for the most number of matrices, METIS is more successful in providing larger percentage of speedup for a smaller set of matrices. This is consistent with our observations in Section~\ref{sec:LB}, where we see that METIS improves both data movement and load imbalance in parallel executions, while RCM is only effective in reducing data movement volume. 

\subsection{\textbf{Pairwise Comparison of Reorderings}}\label{subsubsec:rq1.3}
Comparing reordering methods against one another can further demystify their complex effect in real-world use cases. One important information that we still cannot retrieve from previously shown data in regard to this question is the relative performance of reordering schemes on a matrix by matrix basis. For example, the speedup profile bars for many reordering schemes look quite similar in Figure~\ref{fig:RQ1_perf_profiles}. In this sub-section, we report on pair-wise comparisons across the schemes.

Figure~\ref{fig:winrate_heatmap} presents the fraction of matrices for which the reordering corresponding to the row-label outperforms the reordering scheme corresponding to the column label.

According to this comparison, RCM is the best reordering scheme that achieves better GFLOPs performance compared to all other reordering schemes across all machines except for parallel execution on \textit{Intel-Desktop} where METIS scores a higher win rate. 



\section{Are Reordering Schemes Consistent Across Machines?}
Another important angle in analyzing the effectiveness of reordering schemes is the consistency of their effect. A natural question here is: If a reordering method is successful in yielding higher performance for a chunk of matrices in one machine with a specific architecture, does it hold its promise of achieving a speedup on other machines and architectures?
We answer this question by looking into the matrices that achieve a speedup $>\tau$ for $\tau \in {1.1, 1.25, 1.5, 2}$ on at least one machine -- we define the set of these matrices as Consistency Candidates Set, $CCS$. Then we count how many matrices show inconsistent performance change through the given reordering method by incurring a slowdown on at least one machine -- we define the set of these matrices as Inconsistent Set, $IS$. Our consistency ratio, then, is calculated as follows:

\begin{equation}
    Consistent\% = 1 - \frac{|{IS}|}{|{CCS}|}
\end{equation}
\noindent This consistency ratio allows us to assess if there is a threshold $\tau$ for which the user could expect at least some speedup on all machines if a given speedup is seen on any  machine. If this holds, it would be possible to evaluate a matrix locally on a single machine to determine if the reordering is likely to be superior. 

\begin{figure*}[t]
  \centering
  \includegraphics[width=0.65\textwidth]{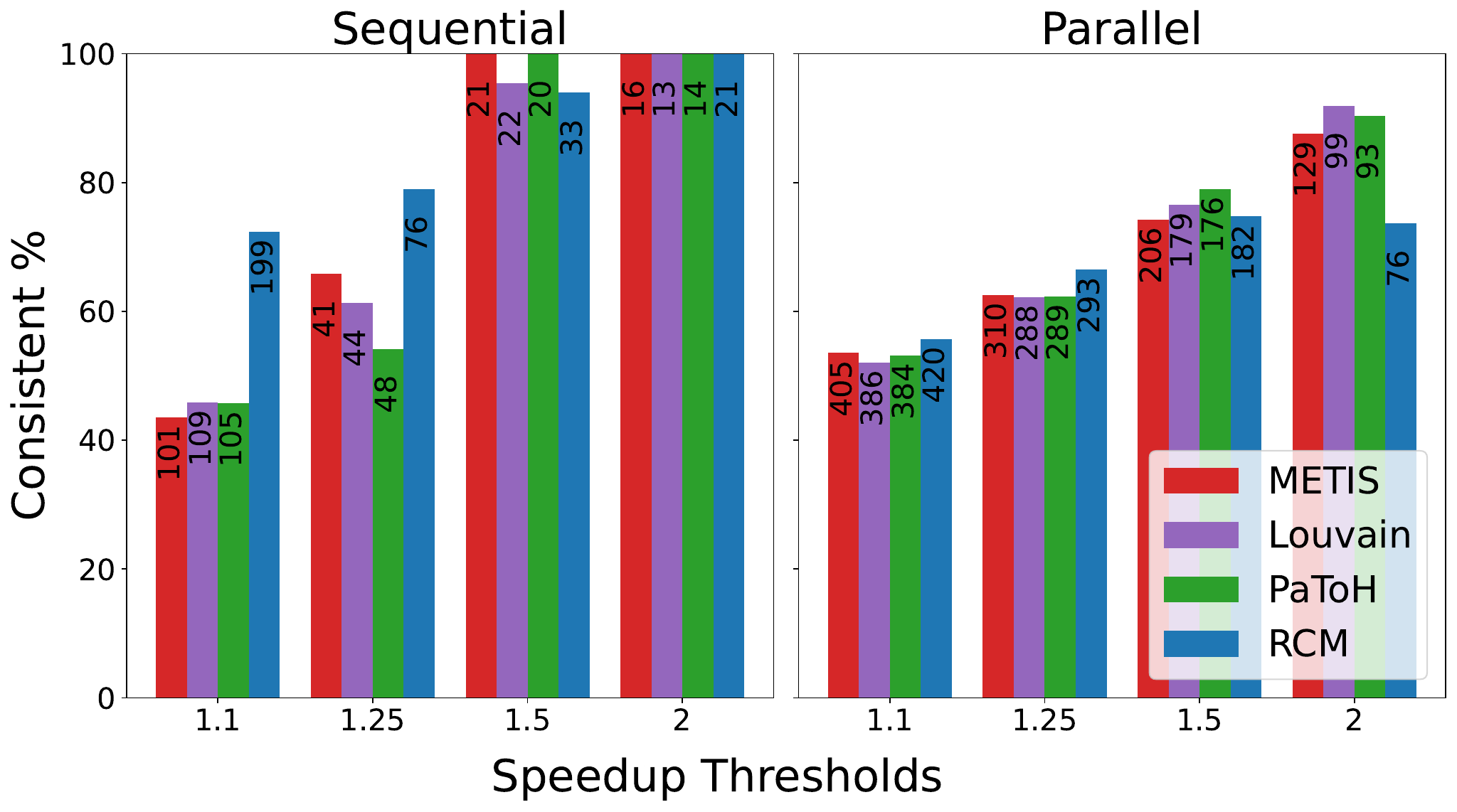}
  \caption{Reordering consistency across machines. The number on each bar represent the number of candidate matrices for that $\tau$ across all machines.}
  \label{fig:RQ3_consistency_machines}
\end{figure*}

Figure~\ref{fig:RQ3_consistency_machines} shows the $Consistent\%$ in a grouped bar plot for different reordering schemes for various threshold values ($\tau$). As expected, by increasing the speedup threshold $\tau$, the $Consistency\%$ increases; this shows that matrices that achieve a big speedup on one machine do not tend to slow down on others.
In the sequential case, RCM leads by being the most consistent scheme across thresholds. However, in the parallel case, there is no clear win for any of the schemes. The consistency ratio ranges from $\approx57$ to $\approx82$ across different thresholds.
Even with a $\tau$ of 2, in the parallel executions, about $\approx20$ of candidates see a slowdown on at least one machine. Reordering for parallel SpMV is considerably machine and architecture dependent.

\section{Impact of Matrix Reordering on Load Balance}
\label{sec:reorder}
In single-core execution, performance differences caused by matrix reordering stem from data movement and caching effects. In parallel execution, however, reordering influences both data access patterns and workload distribution among processing cores. For example, a baseline Compressed Sparse Row (CSR) implementation of Sparse Matrix-Vector Multiplication (SpMV) assigns each core a set of rows, with the workload proportional to the number of non-zero elements per row. Matrix reordering redistributes non-zeros within rows and across row bands, inherently altering the workload balance across cores.
In static scheduling, the workload (e.g., rows of a sparse matrix in an SpMV computation) is pre-assigned to processors before execution begins (see Section \ref{sec:background}). This assignment remains fixed throughout the runtime. While this approach minimizes scheduling overhead, ensures predictable execution, and has been shown to be the best OpenMP schedule for a CSR SpMV (see Figure \ref{fig:schedules}); it has a significant drawback: load imbalance.

\subsection{Non-Zero Load Imbalance}
The number of arithmetic operations and the associated data volume are directly proportional to the number of nonzero entries (nnz) processed. Consequently, the nnz assigned to each processor serve as an effective indicator of workload imbalance. For a simple Compressed Sparse Row (CSR)-based sparse matrix-vector multiplication (SpMV), rows are distributed among processors. Under a static OpenMP scheduling approach, rows are allocated in a block cyclic manner. Specifically, in the default static schedule, each processor is assigned a contiguous block of rows, ensuring each processor processes an equal number of rows consecutively.

We define nnz load imbalance as:
\[
\begin{aligned}
\text{Load Imbalance} &= \frac{\text{max\_load}}{\text{fair\_load}} 
\quad \quad
\text{fair\_load} = \frac{\text{total\_nnz}}{\text{\#processors}}
\end{aligned}
\]
\noindent where $\text{max\_load}$ is the maximum number of nonzeros (NNZs) assigned to any processor, and $\text{fair\_load}$ is the ideal equal distribution of NNZs across processors, computed as the ratio of $\text{total\_nnz}$ to the number of processors.

\begin{figure}[ht]
  \centering
  \includegraphics[width=0.48\textwidth]{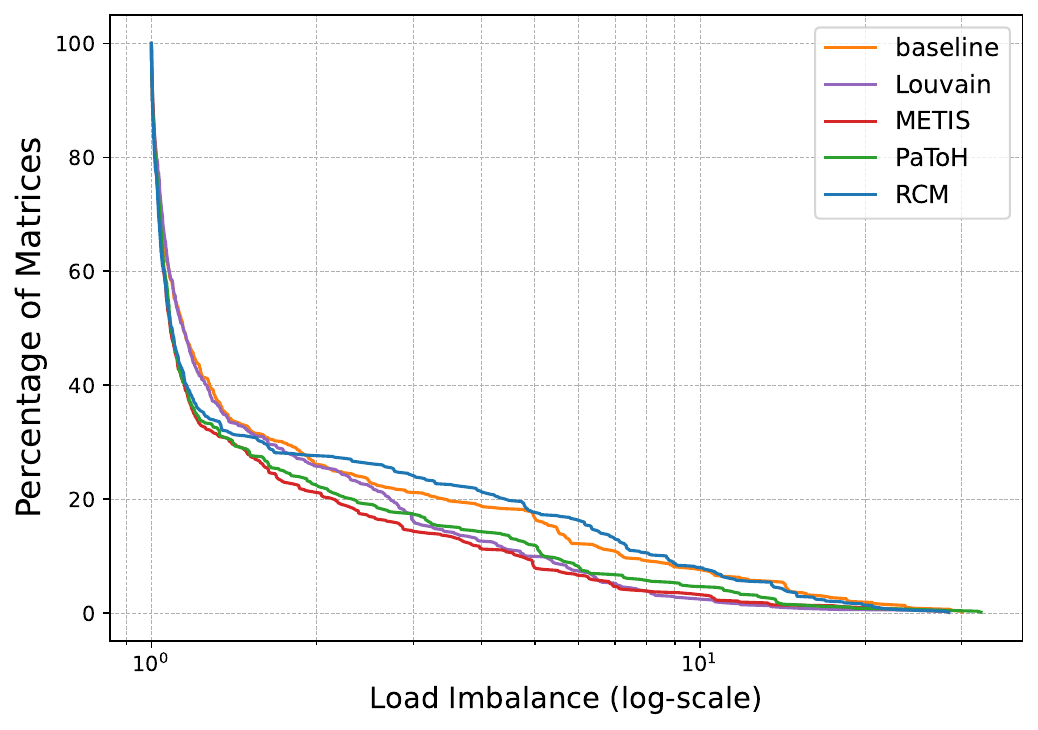}
  \caption{Visualization of the Impact of Reordering Techniques on NNZ Load Imbalance for Static Scheduling Across 64 Threads.}
  \label{fig:LI_all}
\end{figure}
\begin{figure*}[ht]
  \centering
  \begin{minipage}[t]{0.48\textwidth}
    \centering
    \includegraphics[width=\linewidth]{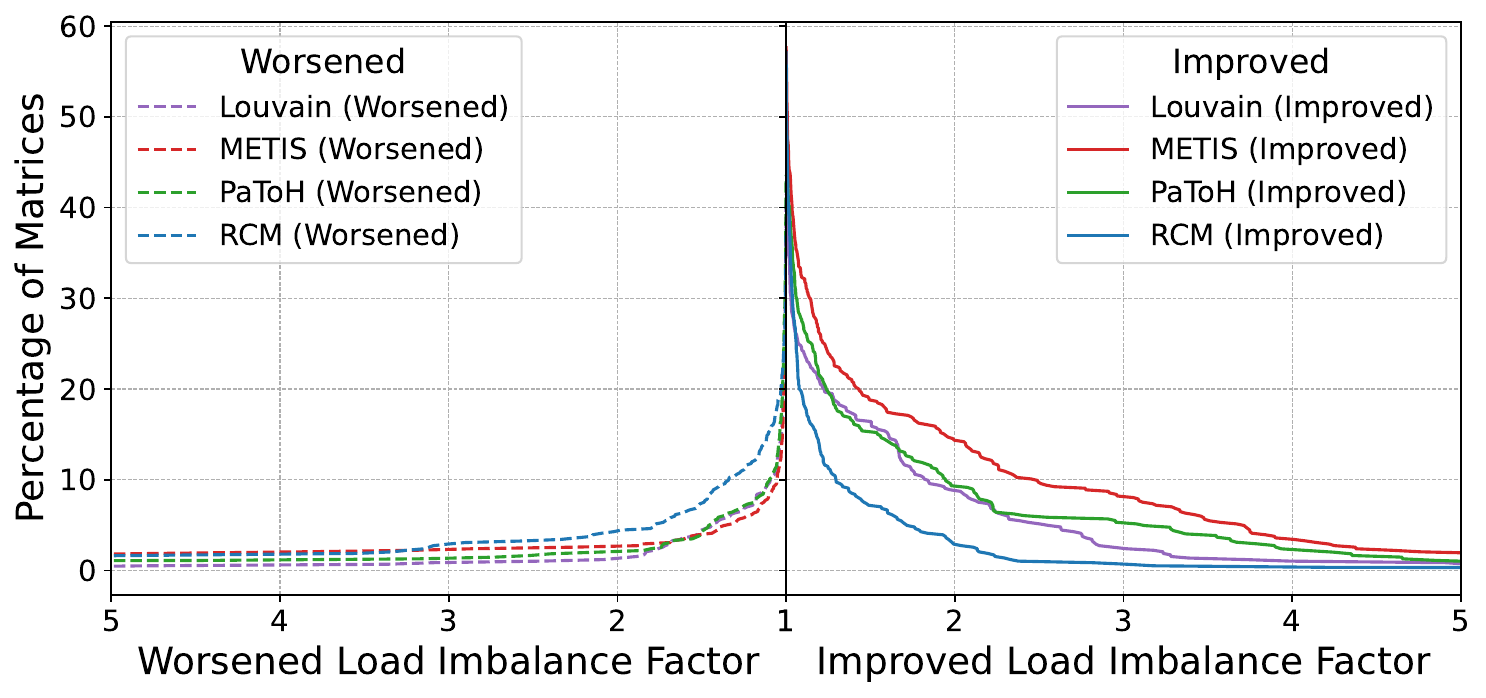}
  \end{minipage}
  \hfill
  \begin{minipage}[t]{0.48\textwidth}
    \centering
    \includegraphics[width=\linewidth]{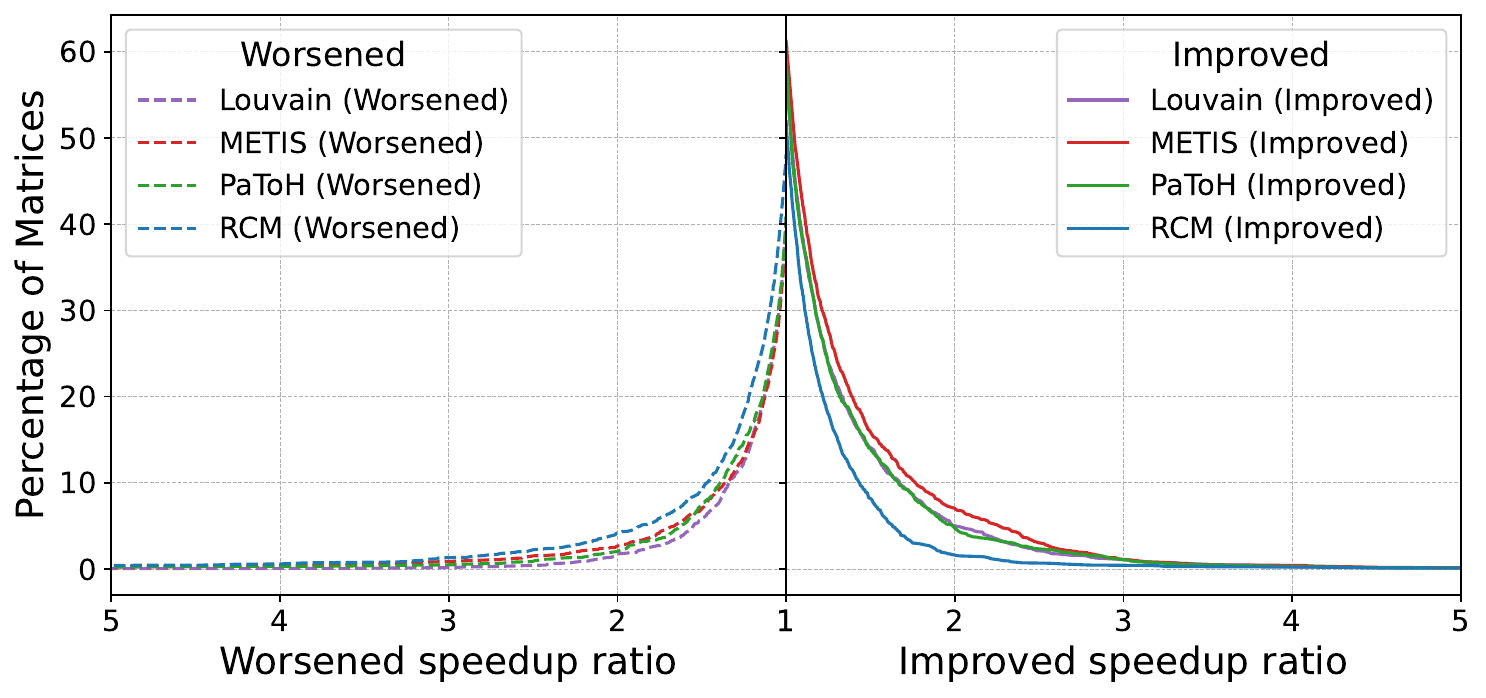}
  \end{minipage}
  \caption{(Left): Impact of Reordering Schemes on Theoretical Load Imbalance Relative to the Baseline Matrix Order. (Right): Speedup ratio in parallel over sequential execution}
  \label{fig:LI_combined}
\end{figure*}

Figure~\ref{fig:LI_all} illustrates the non-zero (nnz) load imbalance associated with the row block assignment employed by the OpenMP static scheduler across 64 threads. This visualization effectively demonstrates the impact of various reorderings on the static load distribution relative to the baseline configuration. Among the tested methods, \texttt{METIS} delivers the most substantial improvement, markedly reducing the imbalance compared to the baseline. In contrast, \texttt{RCM} either fails to enhance or even exacerbates the load imbalance relative to the baseline.
%
To further illustrate the impact of reordering schemes on load imbalance, Figure~\ref{fig:LI_combined}~(Left) presents the relative changes in load imbalance compared to the baseline configuration. Matrices that demonstrate improved load balance are represented as $X/Baseline$, while those showing a deterioration are displayed as $Baseline/X$. The results indicate that \texttt{METIS} achieves the most significant improvement in load balance. Similarly, both \texttt{PaToH} and \texttt{Louvain} exhibit comparable performance, with \texttt{PaToH} displaying a slight advantage. Conversely, \texttt{RCM} shows the least improvement and results in the largest negative impact on load balance relative to the baseline. These findings are consistent with prior observations (e.g., Figure~\ref{fig:RQ1_perf_profiles}), where RCM exhibited superior performance in sequential execution due to load balance being less critical in that context. However, in parallel execution, \texttt{METIS} achieves greater speedups for certain matrices by optimizing workload distribution across processors, thereby effectively improving load balance.

Fig.~\ref{fig:LI_combined}~(Right) presents cumulative distribution of the following ratio computed for each matrix:
\[
\frac{GFLOPS_{parallel}^{reordered}/GFLOPS_{parallel}^{original}}
     {GFLOPS_{sequential}^{reordered}/GFLOPS_{sequential}^{original}}
\]

This ratio represents the extent by which the speedup in the parallel case from reordering exceeds the speedup in the sequential case from reordering. It may be seen that the relative order of the CDF curves for the reordering schemes is the same as that seen in Fig.~\ref{fig:LI_combined}~(Left), confirming that the relative improvements for some reordering schemes like \texttt{METIS} over \texttt{RCM} is indeed attributable to the improvements in load-balancing.




\subsection{Empirical Effects of Reordering on Load Balancing}
\label{sec:LB}
To isolate the non-load balancing effects of matrix reordering on sparse matrix-vector multiplication (SpMV), we develop a load-balanced OpenMP implementation. This implementation allows for the analysis of performance differences between the default static schedule and a custom load-balanced schedule. Any observed throughput differences between the static and load-balanced schedules can be directly attributed to the effects of load balancing. Additionally, the results obtained from the load-balanced schedules effectively isolate the impact of data movement, ensuring that load balancing does not confound the analysis.

Listing~\ref{lst:nnzb} is the \texttt{nnz}-balanced scheduling algorithm used for the custom nnz balanced schedule. This approach seeks to equalize the distribution of nonzero entries across processors, directly addressing workload imbalance more effectively than the default static scheduling method.

\begin{FloatLst}
\begin{lstlisting}[language=C++, caption={CSR SpMV with load balancing by nonzero count.}, label={lst:nnzb}]
void csr_mv(A_CSR, m, x, y, rowPanel_start) {
    #pragma omp parallel
    {
        tid = omp_get_thread_num();
        start_row = rowPanel_start[tid];
        end_row = rowPanel_start[tid + 1];

        for (r = start_row; r < end_row; r++) {
           sum = 0;
           for (idx = A_CSR->rowPtr[r]; idx<A_CSR->rowPtr[r+1];idx++)
           {
              col = A_CSR->cols[idx];
              val = A_CSR->values[idx];
              sum += val * x[col];
           }
           y[r] = sum;
        }
    }
}
\end{lstlisting}
\end{FloatLst}

\begin{figure*}[ht]
  \centering
  \includegraphics[width=0.7\textwidth]{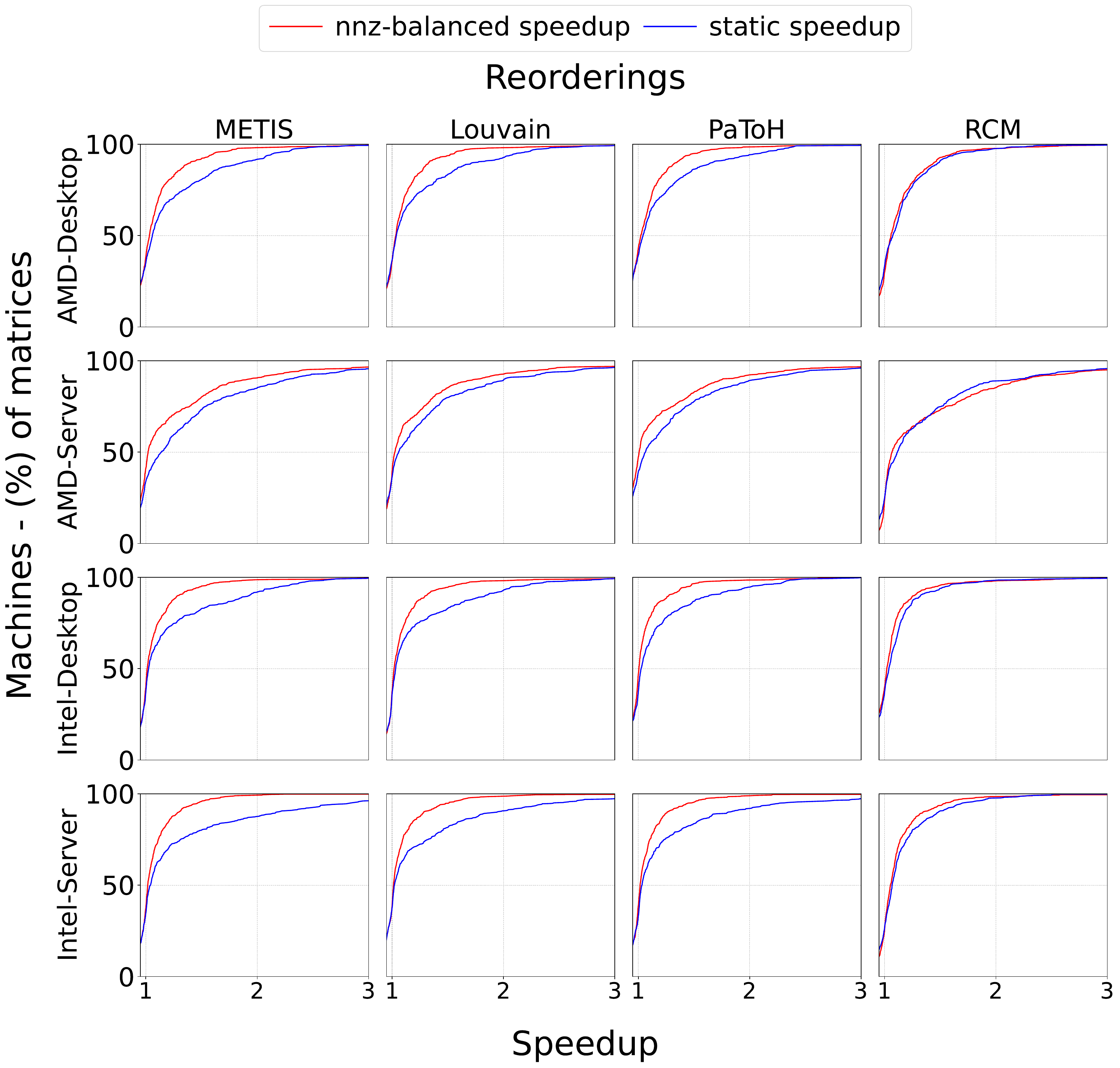}
  \caption{Comparative analysis of speedups between \texttt{nnz}-balanced and static scheduling.}
  \label{fig:LB-empirical}
\end{figure*}

Figure~\ref{fig:LB-empirical} provides a comparative analysis of speedups achieved by the \texttt{nnz}-balanced and static scheduling algorithms across various architectures, employing a reverse CDF to summarize the results for all reordering strategies. For reorderings such as \texttt{METIS}, \texttt{Louvain}, and \texttt{PaToH}, the \texttt{nnz}-balanced curves exhibit a distinct advantage over their static scheduling counterparts. This gap demonstrates the significant contribution of reordering-induced load balance improvements to the performance gains achieved by these methods. However, the \texttt{nnz}-balanced scheduling effectively mitigates the imbalance the effectiveness of the \texttt{nnz}-balanced scheduling effectively mitigates the imbalance, resulting in muted speedups gains in the presence of a load balancing schedule.
Conversely, \texttt{RCM} displays a contrasting behavior, as its \texttt{nnz}-balanced and static scheduling curves are nearly identical. This observation aligns with prior load imbalance analyses, where \texttt{RCM} was found to be ineffective at improving static load imbalance relative to the baseline. Therefore, any performance enhancements resulting from the \texttt{RCM} reordering are likely driven by favorable data access patterns rather than by improved load balancing.


\section{Related Work}\label{sec:relatedwork}
Recent comprehensive studies on comparing reordering schemes on multi-core platforms have shown that graph and hypergraph -based methods are generally superior in terms of achieved performance~\cite{Gao2024ASL, sc23}. There is an important gap in prior work, which is the fact that simple repeated measurements do not simulate or reflect real-world application performance. The presented geomean of achieved GFLOPS from \cite{sc23} lead to finding METIS as the best reordering overall, while this is not completely aligned with our findings through our IOS measurement method.

In \cite{Weiland2012MixedmodeIO}, authors show that the reordering impact on performance in single-core settings is most pronounced with smaller matrices that fit in the cache. Authors in \cite{Azad2016TheRC} discuss the advantages of reordering schemes extending beyond those seen on single-core machines by looking into the reduction in inter-core communication and cache effects. They highlight RCM's capability in improving data locality, matching our experimental observations.

The pivotal role of reordering schemes in sparse linear system solutions by reducing the maximum distance between the main diagonal of the matrix and its non-zero elements is analyzed in~\cite{Li2015AnalysisOA}. Authors in \cite{Liang2024OrthogonalBK} depict how this bandwidth reduction improves the orthogonality between row blocks in the matrix. RCM method -- by concentrating non-zero element around the diagonal -- is found to be the most effective in bandwidth and profile reduction and improving computational efficiency by \cite{Hou2024RCMReverseCO} and \cite{Liang2024OrthogonalBK}.
\section{Discussion and Conclusions}


In this paper, we have conducted a number of experiments to gain insights into the question: {\em Is sparse matrix reordering effective for sparse matrix-vector multiplication?} 

We have addressed this question in the context of shared-memory multicore systems. The overall conclusion is that in contrast to the distributed-memory context, where matrix reordering via graph partitioning techniques is used in practice because of consistent improvements, the same is unfortunately currently not the case for the shared-memory context and more research is needed on this topic. A few key takeaways from our study are summarized below.

{\noindent \bf Measurement Methodology:} 
The typical experimental measurement methodology of making repeated measurements to address run-to-run variability can lead to misleading conclusions compared to what actually happens in a typical usage scenario. We introduced the {\em IOS} methodology of swapping input and output vectors for repeated SpMV, and showed that it better matches SpMV performance within applications like a conjugate gradient sparse solver.

In \S~\ref{subsec:rq1} on pairwise comparison of reordering methods, we observed that RCM outperforms other reordering schemes including METIS; however, this is in contrast with conclusions that would have been drawn if we had -- similar to results from \cite{sc23} -- used the standard methodology of repeated measurements without any input-output swap. To further elaborate, we present Table~\ref{tab:disc_gflops_raxios} here using the two measurement methods that were introduced in \S~\ref{subsec:rq0} (IOS and RAX) in comparison with the real-world application, Conjugate Gradient (CG).
Our conclusion from \S~\ref{subsubsec:rq1.3} that RCM is the best reordering holds for both IOS and CG while looking at RAX measurement, an opposite conclusion would be that METIS is the better scheme on three out of the four platforms. 

\begin{figure}[h]
  \centering
  \includegraphics[width=0.45\textwidth]{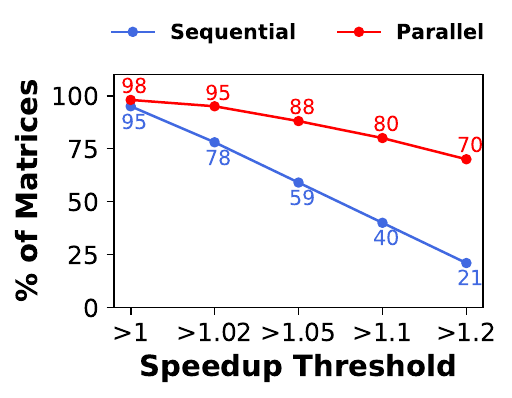}
\caption{Percentage of matrices achieving speedup on at least one reordering/machine configuration.}
\label{fig:speedup_any_any}
\end{figure}

{\noindent \bf Surprising extent of slowdown from reordering:} A somewhat surprising observation is that all reordering schemes actually cause performance degradation for a significant fraction of matrices. For the sequential case, except for RCM, all reordering schemes actually cause slowdown after reordering for more than 50\% of matrices! However, if we ask what fraction of the matrices can achieve some performance improvement on at least one target machine with at least one reordering scheme, the results are encouraging, as seen in Fig.~\ref{fig:speedup_any_any}. For the parallel case, 98\% of matrices achieve a speedup with some reordering scheme on some platform. Is it something in the inherent structure of these matrices? Can reordering schemes be devised that are robust in this regard and do not cause performance deterioration? These are research questions for the future.

{\noindent \bf Lack of consistency across machines:} A question of interest is: {\em Can a reordered matrix be used effectively across machines?} If this is the case, a one-time reordering could be used and the matrix stored in that format (e.g., in the SuiteSparse collection). Data sets are often used in many analyses over time and therefore this question is of interest. Our conclusion (Sec. 5) is that unfortunately there is lack of consistency across machines for a significant fraction of matrices - speedup from the reordered matrix on one machine but a slowdown on another. This leads to the question: Can reordering schemes be devised that exhibit a higher level of consistency?

{\noindent \bf Impact on Load Imbalance:} 
Section \S~\ref{sec:reorder} shows that load balancing is an important aspect of reordering performance. METIS has the largest positive impact on load balance, while RCM does not improve load balance over the baseline. Despite its favorable speedup characteristics, RCM exhibits the lowest improvement in load balance among the evaluated reordering strategies. Consequently, the speedup gains observed with RCM can be attributed primarily to data movement and locality optimizations rather than enhanced load balancing. 
In parallel execution, METIS demonstrates better speedups for certain matrices, which can be attributed to its ability to achieve better load balance compared to RCM. An open research topic is to develop reordering schemes that are designed to improve load balancing while also optimizing data locality. 

\begin{table}[h]
\centering
\caption{RCM versus METIS wins (w) and losses (l) under
IOS, RAX, and CG}
\setlength{\tabcolsep}{7pt}
\begin{tabular}{cx{5mm}x{5mm}|x{5mm}x{5mm}|x{5mm}x{5mm}}
\toprule
\multicolumn{1}{c}{\multirow{2}{*}{}} &  \multicolumn{2}{c}{IOS} & \multicolumn{2}{c}{CG} & \multicolumn{2}{c}{RAX} \\
\cmidrule(lr){2-3} \cmidrule(lr){4-5} \cmidrule(lr){6-7}
\textit{} & \textit{w} & \textit{l} & \textit{w} & \textit{l} & \textit{w} & \textit{l} \\
\cmidrule(lr){2-2} \cmidrule(lr){3-3} \cmidrule(lr){4-4} \cmidrule(lr){5-5} \cmidrule(lr){6-6} \cmidrule(lr){7-7}
AMD-Desktop & \textbf{323} & 264 & \textbf{323} & 264 & 257 & \textbf{330} \\
AMD-Server & \textbf{400} & 187 & \textbf{383} & 204 & 277 & \textbf{310} \\
Intel-Desktop & 271 & \textbf{316} & 246 & \textbf{341} & 242 & \textbf{345} \\
Intel-Server & \textbf{357} & 230 & \textbf{359} & 228 & \textbf{308} & 279 \\
\bottomrule
\end{tabular}
\label{tab:disc_gflops_raxios}
\end{table}

\begin{acks}
This work was supported in part by the U.S. National Science Foundation through award 2009007.
\end{acks}

\bibliographystyle{ACM-Reference-Format}
\bibliography{main.bbl}
\end{document}